%% file: main.tex
\documentclass[10pt,twocolumn,letterpaper]{article}

\usepackage{cvpr}              % To produce the CAMERA-
\input{preamble}

\definecolor{cvprblue}{rgb}{0.21,0.49,0.74}
\usepackage[pagebackref,breaklinks,colorlinks,citecolor=cvprblue]{hyperref}

\def\paperID{*****} % *** Enter the Paper ID here
\def\confName{CVPR}
\def\confYear{2024}

\title{NewsCaption: Named-Entity aware Captioning for Out-of-Context Media}

\author{Anurag Singh
$\quad$
Shivangi Aneja
\vspace{0.3em}\\
{\normalsize Technical University of Munich} \quad
}

\begin{document}
\maketitle
%%%%%%%%% ABSTRACT
\input{tex_code/abstract}
\input{tex_code/introduction}
\input{tex_code/relatedworks}
\input{tex_code/method}
\input{tex_code/implementation_details}
\input{tex_code/experiments}

\input{tex_code/human_evaluation}
\input{tex_code/conclusion_ack}
\input{tex_code/ethics}

{
    \small
    \bibliographystyle{ieeenat_fullname}
    \bibliography{main}
}


\end{document}

%% file: preamble.tex
\usepackage[dvipsnames]{xcolor}

%% file: tex_code/abstract.tex
\begin{abstract}
With the increasing influence of social media, online misinformation has grown to become a societal issue. 
%
%To spread misinformation, adversaries rely on several methods ranging from realistic-looking deepfakes to less compute-intensive methods like cheapfakes.
%
The motivation for our work comes from the threat caused by cheapfakes, where an unaltered image is described using a news caption in a new but false-context. 
%This false-context can be interpreted as an out-of-context caption to spread misinformation.
%
The main challenge in detecting such out-of-context multimedia is the unavailability of large-scale datasets.
%due to annotation of out-of-context multimedia being a cumbersome process. 
%
Several detection methods employ randomly selected captions to generate out-of-context training inputs. However, these randomly matched captions are not truly representative of out-of-context scenarios due to inconsistencies between the image description and the matched caption.
We aim to address these limitations by introducing a novel task of out-of-context caption generation. 
In this work, we propose a new method that generates a realistic out-of-context caption given visual and textual context. We also demonstrate that the semantics of the generated captions can be controlled using the textual context. We also evaluate our method against several baselines and our method improves over the image captioning baseline by 6.2\% BLUE-4, 2.96\% CiDEr, 11.5\% ROUGE, and 7.3\% METEOR
 \end{abstract}

%% file: tex_code/introduction.tex
\section{Introduction}

% Motivation: Why are we working on this problem
% Current trends: What other papers have done to solve the problem
% Limitation of current methods which we fixed
% Our contribution
With recent improvements in several computer vision tasks, manipulation of media has become more realistic threat~\cite{korshunov2018deepfakes,perov2020deepfacelab}. Therefore, many works have recently focused on forensic methods for detection of such deepfakes-based media manipulations~\cite{agarwal2019protecting,afchar2018mesonet,verdoliva2020media,rossler2019faceforensics++,nguyen2019capsule,li2018exposing,yang2019exposing,zhou2017two,aneja2020generalized,cozzolino2021id} 
However, one of the most prevalent ways to spread misinformation to date are cheapfakes, requiring very little to no computing resources compared to deepfakes~\cite{fazio2020out}. 
In this work, we focus on a specific type of cheapfakes, where an unaltered image is used out-of-context with a fake caption to spread misinformation. There has been some recent progress in the detection of out-of-context misinformation~\cite{cosmos,cheapfake_multimodal,akgul2021cosmos,jaiswalmeim,sabirmeir} 
Most of the detection works also propose a method to create these out-of-context captions for the training of detection models, often by selecting other captions at random~\cite{muller2020multimodal,akgul2021cosmos,la2022combination}. 
Others use hypernamed text or similarity methods to select out-of-context captions from a reference set~\cite{jaiswalmeim,sabirmeir}. A study in ~\cite{newsclippings} illustrates that these text manipulations lead to linguistic biases which can be detected without image inputs.\\~\\
To address these challenges, we propose the task of out-of-context caption generation i.e. generating an out-of-context caption for an image given some contextual input. Our task is different from the task of image captioning since unlike image captions news captions are not described by the images alone. To address this, we condition our out-of-context caption on conditional tokens also. The captions generated by our method can by used for downstream tasks such as helping improve out-of-context detection models. The proposed model can also be fine-tuned for the downstream task as a plugged in module. This approach significantly differentiates our method from state-of-the-art methods for generating out-of-context multimedia~\cite{newsclippings} which focus on matching existing unmanipulated text and images in the dataset to obtain out-of-context multimedia.\\~\\
Semantics matching-based methods match a query to a caption in their reference set~\cite{CLIP,ma2022ei,newsclippings}. Therefore, they cannot create captions for unseen named entities outside their reference set. We address this issue using a byte-pair encoding-based captioning module that can process out-of-vocabulary tokens. Additionally, matching-based methods involve some form of linear search within the dataset to match an image to a semantically similar out-of-context caption. Search makes these methods computationally expensive in practice which are also addressed by generating captions instead. 
In summary, our main contributions are as follows:
\begin{itemize}
    \item We propose a new method to generate realistic out-of-context captions given an input image and some conditional word tokens as contextual input.
    \item  We demonstrate that it is possible to control the semantics and context of the generated captions using conditional word tokens. 
   \item Our method addresses the limitations of the previous matching-based methods by generating captions for out-of-vocabulary conditional word tokens.
\end{itemize}

%% file: tex_code/relatedworks.tex
\section{Related Work}

\subsection{Misinformation Detection}
In recent years, models to detect cheapfakes have started to gain attention~\cite{sabirmeir,cosmos,cheapfake_multimodal,la2022combination,akgul2021cosmos}. In ~\cite{cosmos}, authors propose self-supervised training method for out-of-context caption detection. Their model utilizes visual grounding of an image objects in the associated captions and the similarity in contexts of different captions for self supervision. The triplets where captions are grounded in same objects but have different contexts are marked as out-of-context pairs. They also introduce a new large-scale dataset for out-of-context news detection containing news examples collected from different news outlets. Each news example has an image and corresponding news caption. Most of the collected data does not have context annotations. FakenewsNet~\cite{shu2020fakenewsnet} and Fakeddit~\cite{nakamura2019r} focuses on detection of human-made fake news. 

While these works have important real-world use cases for the task of out-of-context multimedia, our focus is on exploring a targeted out-of-context caption where given an image and conditional word tokens, we can automatically generate 
a new out-of-context caption. Our method can address the limitations of randomly selecting out-of-context captions from the dataset by generating a targeted out-of-context caption for the example.

\subsection{Generating Out-of-Context Multimedia}
For generating out-of-context multimedia, matching out-of-context captions in the reference set to corresponding query images and vice versa is done~\cite{newsclippings}~\cite{jaiswalmeim}~\cite{sabirmeir}.
Matching is performed by building a large-scale dataset that acts as a reference set. We try to review several of these datasets in detail. 
The authors propose MAIM~\cite{jaiswalmeim} and MEIR~\cite{sabirmeir} datasets for identifying image re-purposing or discrepancies in the joint semantics of image and text. MAIM takes random captions of other images to create false image caption pairs. MEIR tries to select captions by swapping named entities of people, organizations, and locations in the caption to create false image caption pairs. Recent works ~\cite{muller2020multimodal}~\cite{newsclippings} have released large scale datasets for generating out-of-context multimedia. Authors in ~\cite{muller2020multimodal} release a dataset that has swapped named entities of people, location, and events with other random entities. Authors in ~\cite{newsclippings} argue that this handcrafting from hyper-named text to create out-of-context media generates linguistic biases that are easy to identify. To overcome this limitation ~\cite{newsclippings} proposes a clip-based matching approach for text manipulation. However, matching methods are computationally expensive as obtaining an out-of-context media for a query involves some form of search within the dataset.

These methods assume that for each multimedia image caption pair, another semantically similar pair exists in the reference set whose caption can replace the original caption. Our problem statement is significantly different, as we do not assume the availability of any reference set. Instead, for each image some input named entities act as conditional word tokens to generate the caption.

\subsection{News Captioning} 
The task of news captioning involves generating news captions using the news article and corresponding image as input. In ~\cite{tan2020detecting} authors try to solve the task of fake news captioning by building a model which can replace the entire real news articles with text generated from a large language models like Grover~\cite{zellers2019defending}. However, they do not aim to mismatch the images which are relevant to the news article's content. Analysis of images for this method has limited impact as analyzing articles and captions of the news is the best way to ensure good out-of-context detection performance. Apart from generating fake news articles, several works have focused on automating news captioning of real news~\cite{biten2019good}~\cite{tran2020transform}. Both these methods take a news article and a corresponding image as input to generate a new caption for the article. 

Our task is significantly different from the previous work in news captioning literature since we do not assume any access to complete news articles for our task. At the test time, our model does not need captions, and tokenized named entities can act as input to the model. 
\begin{figure*}[t]
\centering 
\includegraphics[width=0.8075\textwidth]{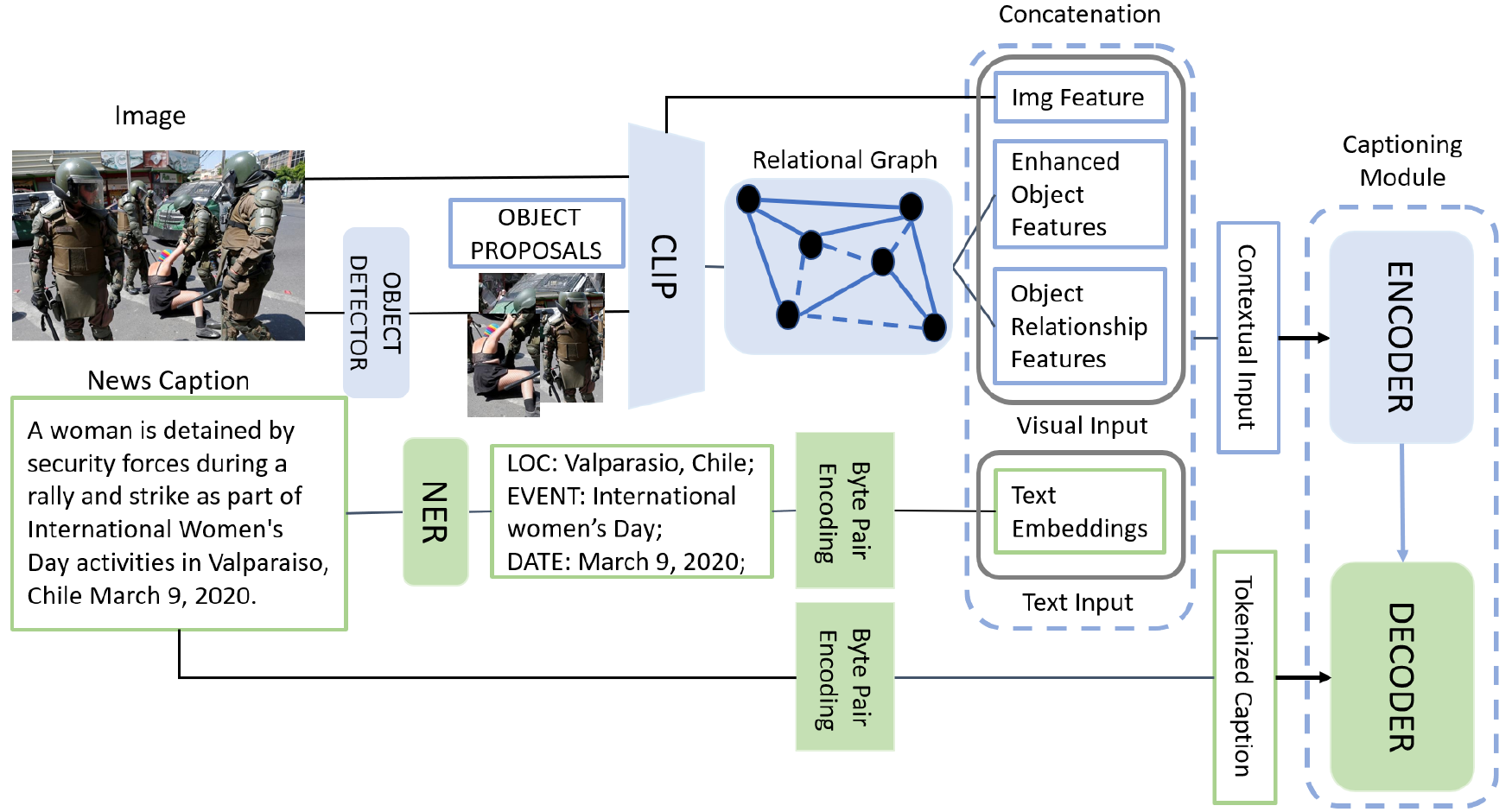}
\label{main}
\caption{The semantic figure describes image and news caption input that are pre-processed using Object-detection, CLIP, and Named Entity Recognition respectively to obtain image and object level features along with a named entity dictionary. The object features are enhanced with a relational graph. The named entities are encoded into text embedding using byte-pair encoding(BPE). The embeddings act as the input to the encoder of the captioning module. Similarly, the original news caption is tokenized using BPE to form input to the decoder during training, and CE loss is used to optimize the model.}
\end{figure*}

%% file: tex_code/method.tex
\section{Method}
We propose an end-to-end architecture to generate out-of-context captions using an input image and conditional tokens. The main components of our model architecture are 1) Named Entity Recognition 2) Feature Extraction \& Detection Backbone 3) Relational Graph 4) Captioning Module.
\subsection{Named Entity Recognition (NER)}
The first step for generating our out-of-context captions is conditioning them on conditional word tokens. During training, we parse the input captions to classify named entities in the text. We use SpaCY NER~\cite{spacy} for processing the captions to identify named entities. This parsing of captions into named entities is only pre-processing of our dataset. We want to highlight that our model does not need the entire caption as input and can work using named entities only. SpaCY classifies input tokens into 18 different categories of named entities. This results in a dictionary $D^{NER}$ with keys as named entity types and tokens of named entities belonging to corresponding types as values.
\subsection{Feature Extraction \& Detection Backbone}
From the image input, we extract image features using a CLIP~\cite{CLIP} visual encode denoted as $\mathcal{I}$. We also detect all probable objects in the given image using an attention-based object detector DETR~\cite{DETR}. The object proposals are extracted from the image using the predicted bounding boxes. These object proposals are encoded into object features using a CLIP visual encoder. 
\subsection{Relational Graph}
\begin{figure}[h]
\includegraphics[width=0.47\textwidth, height=98px]{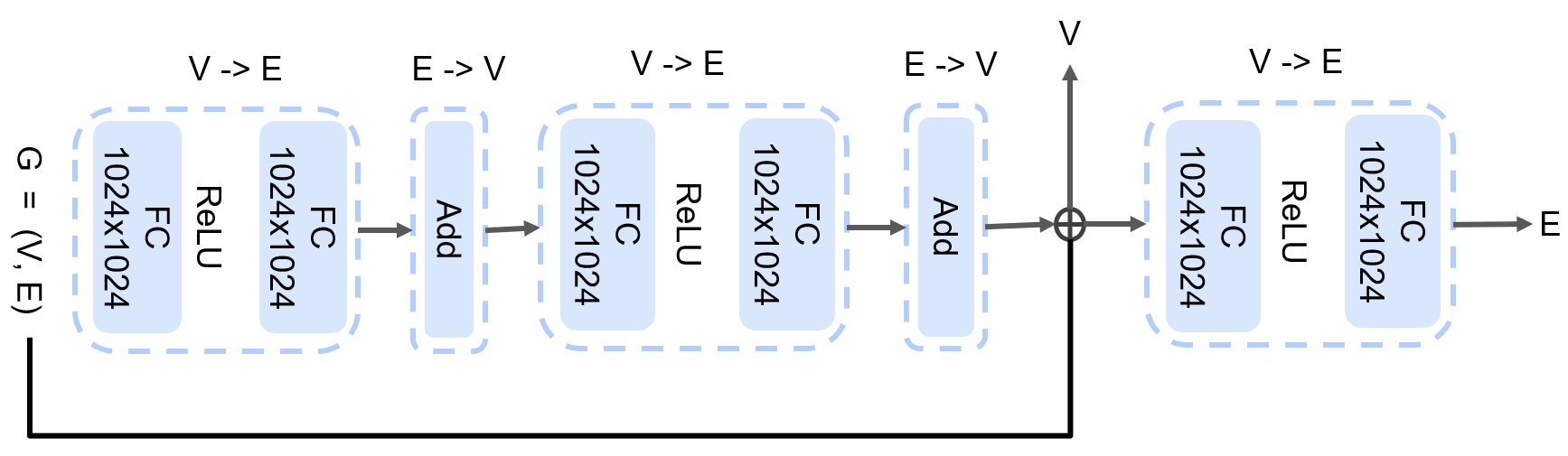}
\label{graph}
\caption{The relational graph module takes CLIP image encoding from object proposals as input that represent nodes in the graph.}
\end{figure}
\begin{figure*}
\centering 
\includegraphics[width=0.9\textwidth]{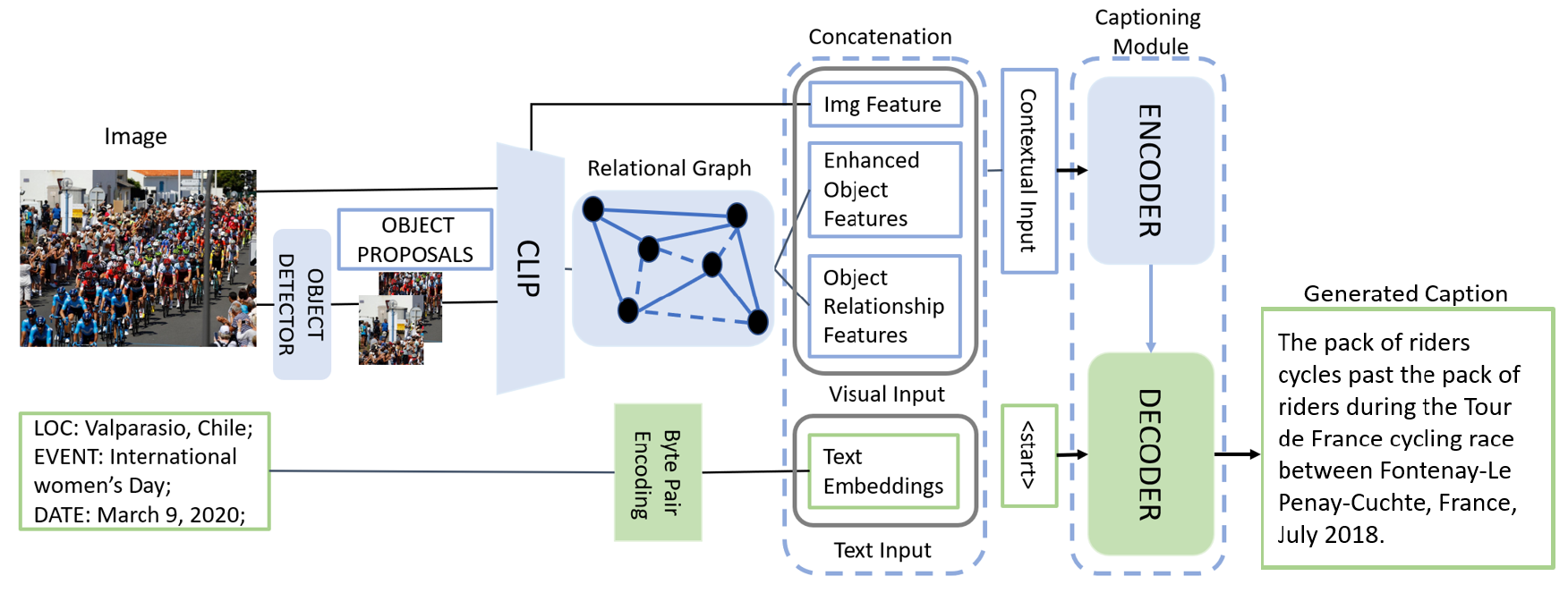}
\label{inference}
\caption{The semantic figure describes the test time image and conditional word token as input to our model. Image is processed using Object-detection and CLIP. Byte pair encoding converts word tokens into text embeddings. These representations form an input to the encoder of the captioning module. We condition the decoder using a start token that denotes the start of a sentence. It then generates a caption in an auto-regressive fashion.}
\end{figure*}
Describing the actions and events in an image often involves understanding the relationship of an object's appearance in context with other objects in the scene. To capture the semantics of such actions and events in the generated out-of-context caption, we employ a graph neural network with message passing to obtain enhanced object features and to extract object relation features from the edges of the relationship graph. We create
a graph $\mathcal{G} = (\mathcal{V}, \mathcal{E})$ where we represent the object features as nodes and the relationship between the objects is represented by edges between the nodes of the graph. We construct the graph where we consider at most $K$ relationships for an object, in other words, each node as $K$ number of edges. We use standard neural message passing ~\cite{gilmer2017neural} where the message passing at graph step $\tau$ is defined as follows:
\begin{equation}
    \mathcal{V} \xrightarrow{} \mathcal{E} : e^{\tau+1}_{i,j} = f^{\tau}([v^{\tau}_{i}, v^{\tau}_{j} - v^{\tau}_{i}]) 
\end{equation}
Where for nodes $i,j$ we denote features at step $\tau$ as $v^{\tau}_{i}\in \mathcal{R}^{1024}$ and $v^{\tau}_{i}\in \mathcal{R}^{1024}$ respectively. $e^{\tau+1}_{i,j}\in \mathcal{R}^{1024}$ represents the edge relationship message between nodes i and j in the next graph step $\tau +1$. The hard brackets [·, ·] represent the concatenation of two vectors, and the function f() is an MLP. The node features are aggregated from the messages post every message passing step as $\mathcal{E} \xrightarrow{} \mathcal{V} : v^{\tau+1}_i =\sum_{k=1}^{K} e^{\tau}_{i,k}$. The node features in the last step are summed to the original features via a skip connection and output as enhanced object features $\mathcal{V}^{\tau}$.  An additional message passing layer after the final layer outputs message $\mathcal{E}^{\tau+1}$ as the object relation features.

\subsection{Captioning Module}
Our captioning module consists of a 3-layer Transformer with an Encoder and Decoder where we feed the visual feature input concatenated to the embeddings of conditional tokens to the Encoder. The byte pair encoded embeddings of the caption tokens are fed as input to the Decoder.  
\subsubsection{Encoder}
We concatenate the image feature vector with the object feature vectors to obtain total visual feature vectors $X^{vis}$.
\begin{equation}
    \mathcal{X}^{vis} = [\mathcal{I}, \mathcal{V}^{\tau}, \mathcal{E}^{\tau+1}]
\end{equation}
These total visual feature vectors $X^{vis}$ acts as input to the encoder. We also consider byte-pair encoding (BPE) to encode the tokens into word embedding. We use pre-trained GPT-2 BPE for our task. The effectiveness of BPE for our task is that it allows us to handle out-of-vocabulary words as compared to word2vec\cite{word2vec} or GLOVE\cite{glove}. We compute the textual features of the full dictionary $\mathcal{D}^{NER}$. We use the BPE of the named entity types to allow our model to understand the context of each conditional token. For each named entity type, we compute the BPE of named entity type and its corresponding tokens which are then concatenated. This concatenation is followed by the embedding of the end token. This allows the model to differentiate between different categories of named entities. The BPE of full dictionary is obtained by concatenating all the vectors. 
\subsubsection{Decoder}
A stack of 3 identical layers constitute the decoder of the captioning module. Similar to architecture in ~\cite{transformers}, decoder receives features from the encoder to compute cross attention. In addition to the byte-pair encoding, we also use a pre-trained position encoding from GPT-2. Position encoding helps in understanding the context of the token in the caption. The final layer generates a prediction over the vocabulary to generate the output token. The decoder masks the embeddings of tokens ahead in the caption while predicting the output of the current token. 

%% file: tex_code/implementation_details.tex
\section{Implementation Details}
We implement our architecture using PyTorch~\cite{paszke2019pytorch} and
train end-to-end using ADAM ~\cite{kingma2014adam} with a learning rate of 1e-3. We train the model for 100 epochs until convergence. We extract a maximum of 10 objects for each image. We pad the object features with 0 for images with fewer objects. 
We consider token length of 20 for our named entity texts. For the examples with less than 20 tokens, we pad the token embeddings with 0. We use the token length in our model to mask attention on the padding.  We truncate input training captions longer than 100 tokens and add $start$ and $end$ tokens from the byte-pair encoding of GPT-2 to indicate the start and end of the caption. 

%% file: tex_code/experiments.tex
\section{Experiments}
\textbf{Dataset:} We use the COSMOS dataset ~\cite{cosmos} of out-of-context news captions which contain images and corresponding news captions from different news outlets. COSMOS consists of 200k images with 450k different captions, corresponding to the true captions from news outlets and also captions corresponding to fake news captions. COSMOS is a dataset designed for self-supervised out-of-context caption detection, therefore, the annotations of true and fake captions are absent. \\~\\
\textbf{Train \& Val Split:} We follow the standard train and val split from COSMOS~\cite{cosmos}, thus use 160K images and corresponding captions. We use the val set to test our model since it has 40k examples in comparison to test set which only contains 1700 examples. We construct a smaller dataset from COSMOS, where we select a single caption corresponding to each image in our training and validation test set.\\~\\
\textbf{Metrics:} To evaluate the quality of generated captions, we used evaluation metrics commonly used for the tasks of image captioning such as BLEU-4 ~\cite{bleu}, CiDEr ~\cite{cider}, METEOR ~\cite{cider} and ROUGE ~\cite{lin2004rouge}. Here BLEU-4 and METEOR allow us to evaluate the effectiveness of the sentence, CiDEr allows us to evaluate the quality of the sentence with respect to the image description, and ROUGE allows us to evaluate the automatic summarization/recall of the generated caption with respect to the caption from which named entities are extracted. \\~\\
\textbf{Baselines:} To evaluate the effectiveness of our method we implement strong baselines using LSTM Seq-2Seq architecture~\cite{vinyals2015show} and with attention~\cite{xu2015show} by modifying them to our task where we use them as our captioning module. For a fair comparison, we use GPT-2 pretrained byte-pair encoding in all the methods. We feed conditional input to Seq-2-Seq-based architecture with our visual features and textual features as input to the encoder which outputs the context embedding that conditions the LSTM decoder to generate out-of-context caption. We also consider 3 layer Transformer baseline with image features and text encoded using BPE but no named entity types and relationship graph as another strong baseline for comparison.  
\subsection{Quantitative Results}
\begin{table}[h]
    \centering
    \scalebox{0.9}{
    \begin{tabular}{|c|c|c|c|c|}
        \hline 
        Method & Bleu-4 & Cider & Rouge & Meteor \\
        \hline
         LSTM &   27.73 & 36.89 & 29.6 & 12.8\\
         LSTM+Attention & 28.9& 40.8& 30.2 &13.4 \\
         Transformer & 31.2 & 47.14 & 33.0 & 15.1 \\
         \textbf{Ours} & \textbf{37.4} & \textbf{50.1} & \textbf{41.5} & \textbf{22.4} \\
         \hline 
    \end{tabular}
    }
    \caption{Comparison of out-of-context caption descriptions using image captioning metrics with modified LSTM Seq-2-Seq architecture with byte-pair encoding, LSTM +attention with byte-pair encoding}
    \label{maincomparison}
\end{table}
We compare our method to baselines on the COSMOS while training on full train set and evaluating on official val set. The results of our method are presented in the Table~\ref{maincomparison}. 
\subsection{Qualitative Results}
\begin{figure*}
    \centering
    \includegraphics[width=0.95\textwidth]{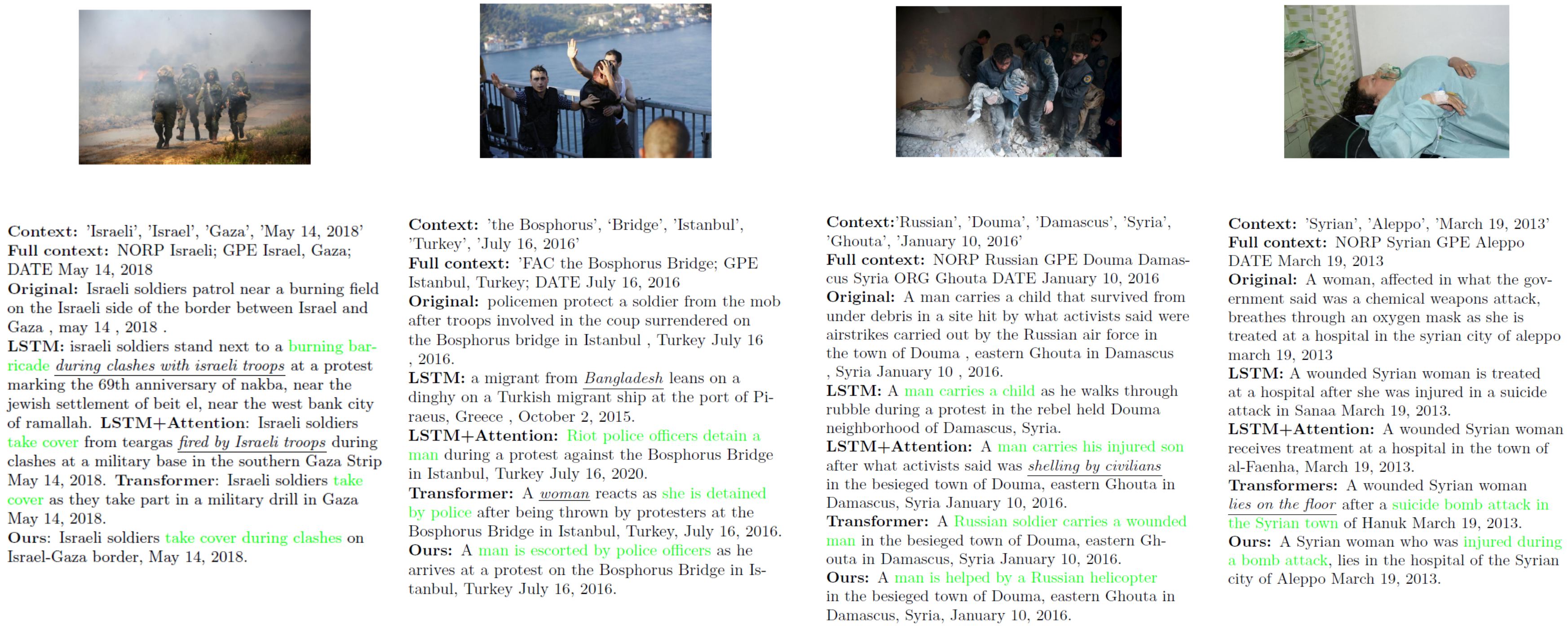}
    \caption{Qualitative comparison of caption generated by different model baselines. The incorrect attributes being included in the caption are highlighted by underlining in the captions. The green highlighting of the text in the caption denotes the semantics which the model understands from the image input.}
    \label{fig:comparison_qualitative}
\end{figure*}

\begin{figure}[htb]
\centering
\includegraphics[width=0.5\textwidth]{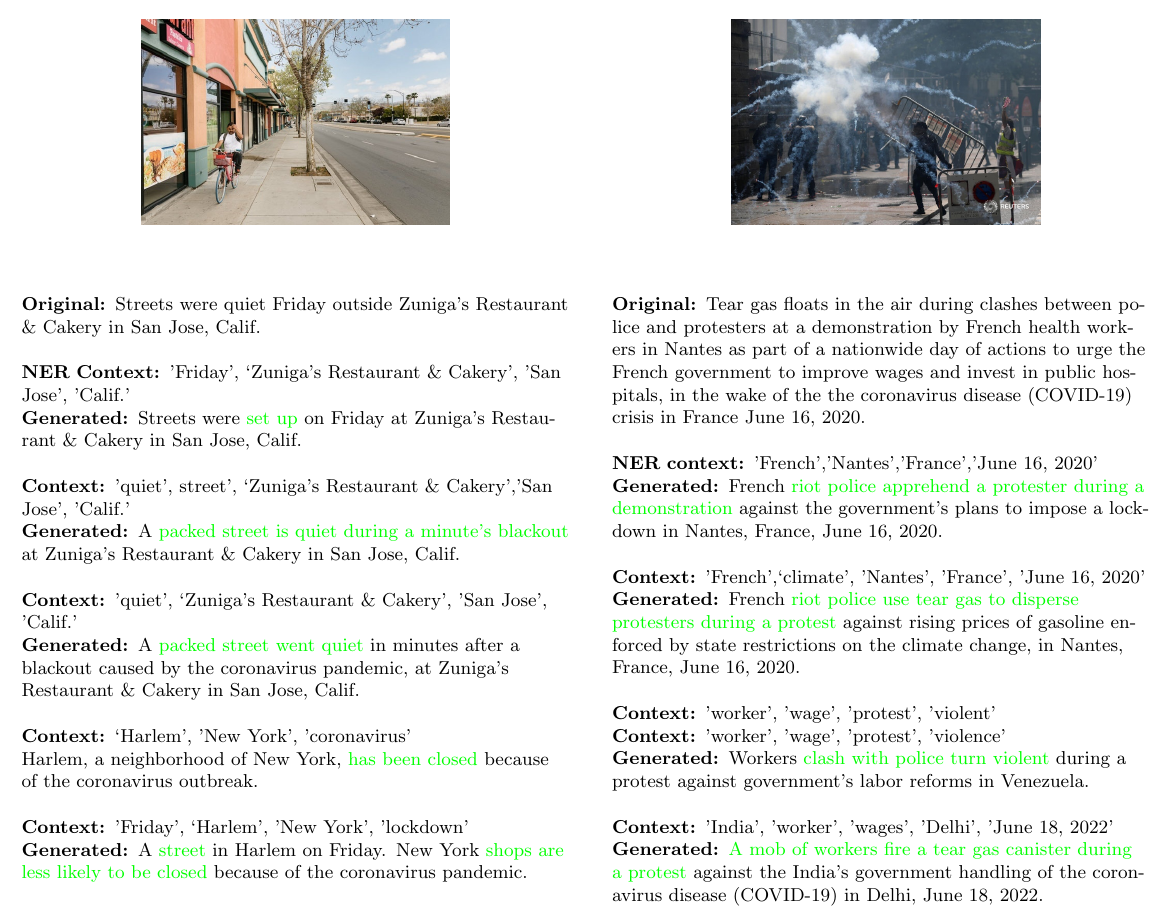}
\caption{Qualitative Comparison of the effect of the conditional word tokens on the semantics of caption generated.
The green highlighted words in the generated caption denote the semantics model implicitly learns from the image input.} \label{fig:playingaround}
\end{figure}

\textbf{Comparison of baselines:} In Figure~\ref{fig:comparison_qualitative} we qualitatively analyze the captions generated by different baseline models. We observe that the use of the Transformer allows for better inclusion of the textual context in the generated caption. In other words, captions are better conditioned on the input text as dates are correct for the transformer model. We observe that using object features and relationship graph allows for better grounding of the caption in the image as transformer model baseline without object features and relation graph mistakes man for women in the caption. The relation graph allows for the model to discover the object relationship between man and the police. This allows the model to include the semantics of man being escorted in the caption.
\\~\\
\textbf{Playing around with the input:} In Figure~\ref {fig:playingaround}, we try to qualitatively analyze captions as we change the conditional word tokens for an unseen image. In this Figure, the NER context denotes the named entities extracted from the corresponding caption in the dataset. We observe for different cases how caption changes as we change the named entities in the context. In the left example of Fig~\ref{fig:playingaround}, we observe that adding the word quiet to the context allows a change in the semantics of the generated caption. Also removing the word street from the context, the generated caption still describes the street from the visual input. A similar observation is also made for the word token Friday. In the right example of the fig~\ref{fig:playingaround}, input from the image allows the model to understand tear gas, protest, and police and use these in the generated caption. Further, changing the context to India and Delhi allows controlling the location of the event. 

\subsection{Ablations}
\begin{figure}[h]
    \centering
    \includegraphics[width=0.45\textwidth]{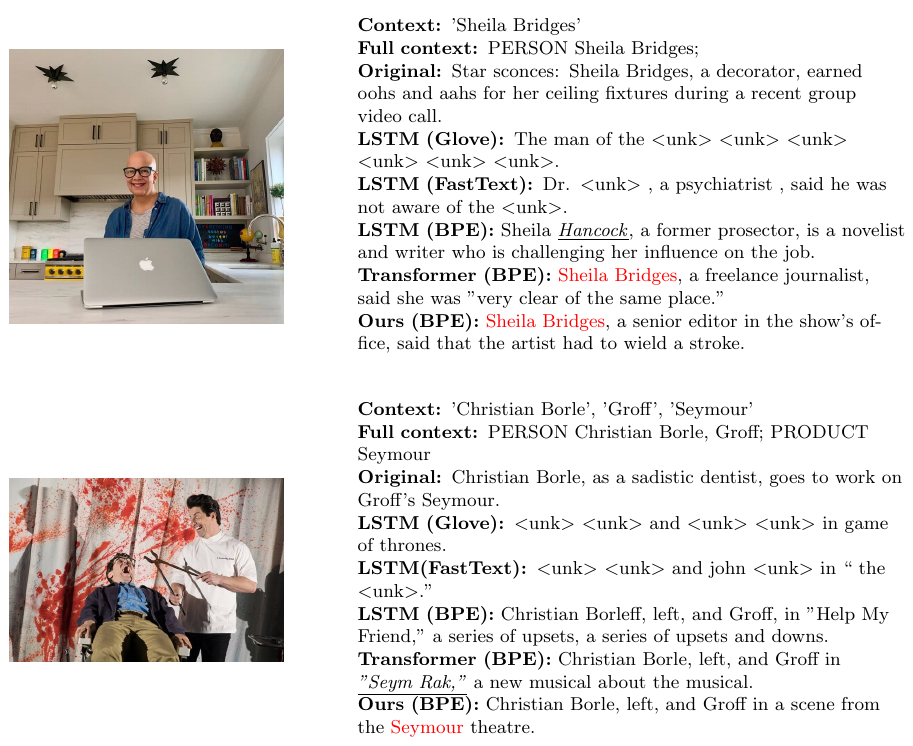}
    \caption{Qualitative comparison of generated caption with different choice of embeddings for LSTM and Transformer captioning backbones. The red highlighted word token in the caption denotes the out of vocabulary word that is best captured by our final model which is tokenized incorrectly in only transformer backbone and is underlined as incorrect.}
    \label{fig:embedding_qualitative}
\end{figure}

\begin{figure}[h]
    \centering
    \includegraphics[width=0.45\textwidth]{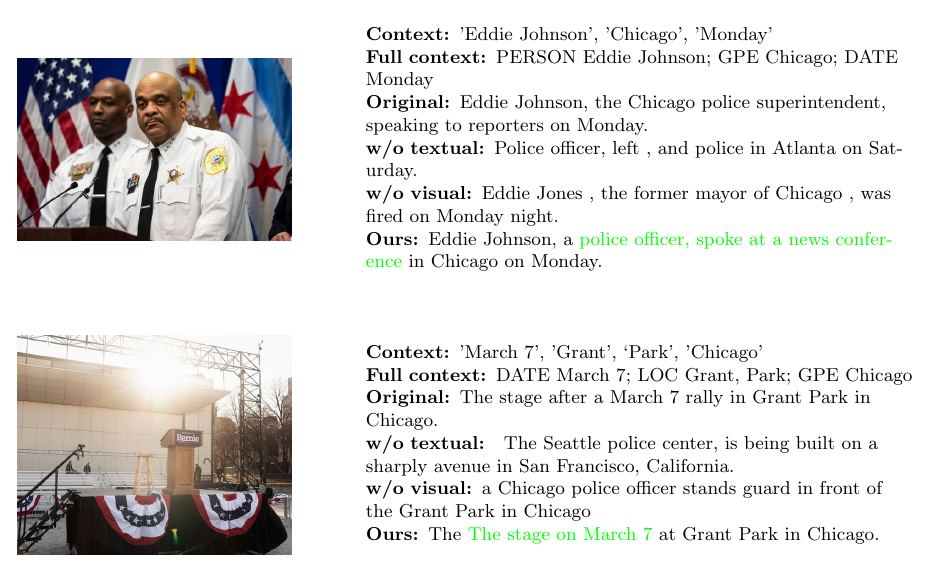}
    \caption{Qualitative comparison of generated caption without different modalities of the contextual input. The green highlighted words in the final caption denotes semantics captured by our final model using both of the modalities in the context.}
    \label{fig:contextualinput_qualitative}
\end{figure}

\begin{figure}
    \centering
    \includegraphics[width=0.45\textwidth]{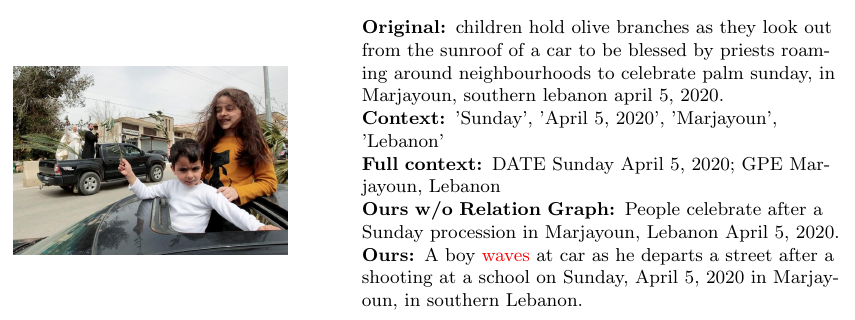}
    \caption{Qualitative comparison of generated caption with and without relational graph. The red highlighted word token in the caption denotes object relationship that is absent from the w/o relationship graph caption.}
    \label{fig:relation_qualitative}
\end{figure}

\textbf{Effect of Embedding?}
We analyze the effect of the choice of embedding for our task for which we take the captioning module to be the LSTM backbone in our experiments. We experiment with different embedding methods in Table~\ref{table_embedding}, where we use Glove, fastText, and pre-trained GPT-2 byte-pair encoding-based embedding. We observe that for our task, BPE outperforms every other embedding since other embeddings simply treat out-of-vocabulary as unknowns. However, BPE tokenizer breaks out-of-vocabulary words into tokens within the learned vocabulary, thus, allowing the model to learn meaningful representations for words it has not seen before. 
\begin{table}[h]
    \centering
    \begin{tabular}{|c|c|c|c|c|}
        \hline 
        Method & Bleu-4 & Cider & Rouge & Meteor \\
        \hline
         Glove & 20.77& 13.92& 22.14 & 8.20\\
         fastText & 21.80 & 15.01& 22.30& 8.50\\
         BPE & \textbf{27.73} & \textbf{36.89} & \textbf{29.6} & \textbf{12.8}\\
         \hline 
    \end{tabular}
    \caption{Ablation for effect of choice of embedding on performance of LSTM baseline\cite{vinyals2015show}. }
    \label{table_embedding}
\end{table}
In Figure~\ref{fig:embedding_qualitative}, we also analyze the qualitative effect of different embeddings on the generated caption. We observe that the use of the Glove and fastText as embedding results in a lot of unknown tokens because these embeddings are trained on different word corpus. Using a self-training vocabulary is also not better than Glove since named entities in the unseen test captions are out-of-vocabulary for the train time corpus as well. We observe that byte-pair encoding works best for our task as it allows us to handle out-of-vocabulary tokens. Using byte-pair encoding generates more meaningful captions compared to Glove and fastText for challenging examples. We underline a limitation of byte-pair encoding in the figure for Transformer backbone caption. Given less context and challenging out-of-vocabulary word, BPE can misconstrue it.
\\~\\
\textbf{Effect of Loss: } We analyze the effect of loss used in our model on the output probabilities over the vocabulary on training. Since COSMOS dataset has different category of frequencies for topics, we compare cross entropy effect over with weighted cross entropy and focal loss. For weighted cross entropy, we compute the weights by computing the frequency of each of the 50k tokens in our byte-pair encoding. There is marginal improvement in the performance on considering the imbalance in the tokens within the loss.
\begin{table}[h]
    \centering
    \begin{tabular}{|c|c|c|c|c|}
        \hline 
        Method & Bleu-4 & Cider & Rouge & Meteor \\
        \hline
         CE & 37.4 & 50.1 & 41.5 & 22.4\\
         Weighted CE & 37.4 & 51.0 & 41.2 & 22.6 \\
         Focal & 37.8 & 51.2 & 41.8 & 22.9\\
         \hline 
    \end{tabular}
    \caption{Comparison of Cross Entropy (CE), Weighted Cross Entropy and Focal loss over output probabilities of decoder.}
\end{table}
\\~\\
\textbf{Ablation on contextual input:}
The contextual input to the encoder in the captioning module consists of two modalities. Firstly, the \textit{visual} input which includes the image feature, enhanced object features and the object relationship features. Secondly, the \textit{textual} input which includes BPE of the conditional word tokens and the named entity types. In the table~\ref{tab:contextual_inputs} we analyse the effect of different modalities on the performance of our model. We observe that both modalities significantly help the performance of our model and the model performance significantly drops with the removal of any one of these modalities. In fig~\ref{fig:contextualinput_qualitative} we demonstrate this qualitatively with the help of an example. In the figure, the removal of textual input from the context limits the model capacity to caption the news with correct named entities. On the other hand, removal of the visual input limits the models capacity to understand the relationships between the input named entities and the generated caption is no longer grounded in the reference images.   
\begin{table}[h]
    \centering
    \scalebox{0.95}{
    \begin{tabular}{|c|c|c|c|c|}
        \hline 
        Method & Bleu-4 & Cider & Rouge & Meteor \\
        \hline
        w/o visual & 28.0 & 43.6 & 30.9 & 13.2\\
        w/o textual & 22.1 & 32.4 & 27.1 & 12.0\\
        \textbf{Ours} & \textbf{37.4} & \textbf{50.1} & \textbf{41.5} & \textbf{22.4}\\
        \hline 
    \end{tabular}
    }
    \caption{Comparison of the effect of visual and textual input on the overall performance of the model}
    \label{tab:contextual_inputs}
\end{table}
\\~\\
\textbf{Are entity types in textual input helpful?}
We analyze the effect of the entity types in the textual input. In Table~\ref{tab:named_entity_types} we consider the ablation of the model without named entity types to with with respect to overall model which is denoted by Named Entity Type + Relational Graph. We observe that removing the named entity types from the text reduced the performance of the model as it removes the input from the model to know how to use the named entities in what context. We also observe that none of the generated captions when including named entity types contain entity types in the text, which underscores that model learns them only as semantics.
\begin{table}[h]
    \centering
    \scalebox{0.9}{
    \begin{tabular}{|c|c|c|c|c|}
        \hline 
        Method & Bleu-4 & Cider & Rouge & Meteor \\
        \hline
        w/o NET &33.1 & 47.0 & 34.4 & 17.6 \\
        NET & 35.2 & 47.8 & 37.5 & 20.1 \\
        NET+Relational Graph & 37.4 & 50.1 & 41.5 & 22.4\\
        \hline 
    \end{tabular}
    }
    \caption{Comparison of effect of removal of named entity types (NET) vs removing the named entity types}
    \label{tab:named_entity_types}
\end{table}
\\~\\
% added human evaluation figure in middle to format the draft better
\begin{figure*}
    \centering
    \includegraphics[width=\textwidth]{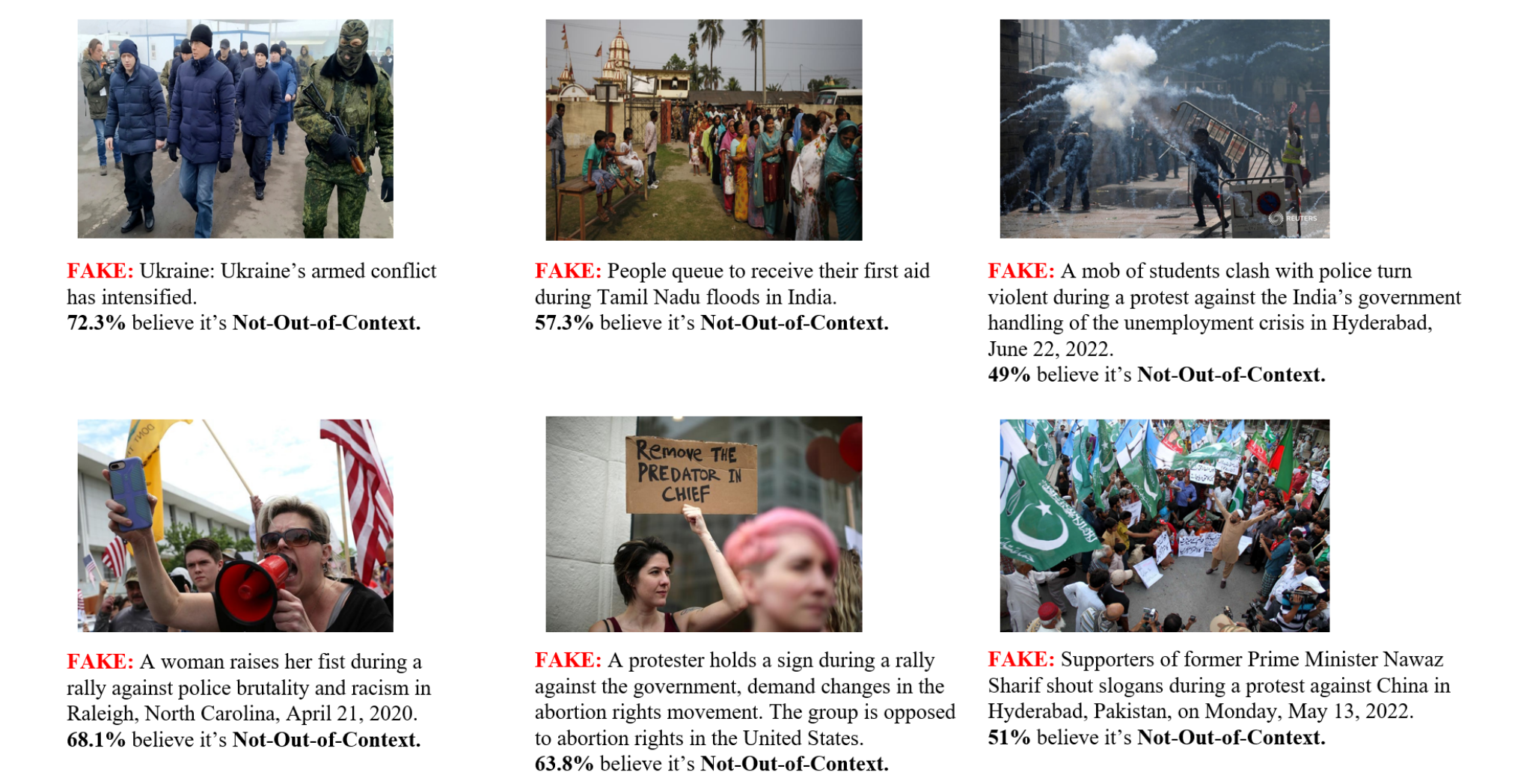}
    \caption{Qualitative comparison of 6 out-of-context news examples from the user study. Where we denote the model generated captions as fake. We also provide a statistic of how many respondents misclassify each news example}
    \label{fig:userstudy_qualitative}
\end{figure*}
\textbf{Is using a relational graph helpful?}
We analyze the effect of relational graph on the performance of our model. In Table~\ref{tab:relation_graph_ablation} we consider the ablations of our model without relational graph and then within the relation graph we selectively remove the enhanced object features and object relationship/edge features. We observe that the enhanced object features alone do not allow the model to understand and incorporate objects in the sentences as incorporating objects in the caption requires context and the needs to have relationship between objects learnt as form of object relationship features that are extracted from the edges of the relation graph.
\begin{table}[h]
    \centering
    \scalebox{0.95}{
    \begin{tabular}{|c|c|c|c|c|}
        \hline 
        Method & Bleu-4 & Cider & Rouge & Meteor \\
        \hline
        w/o relational graph & 35.2 & 47.8 & 37.5 & 20.1 \\
        w/o edge features & 35.3 & 48.0 & 38.2 & 19.0\\
        w/o object features & 36.0 & 48.1 & 39.6 & 20.5 \\
        \textbf{Ours} & \textbf{37.4} & \textbf{50.1} & \textbf{41.5} & \textbf{22.4}\\
        \hline 
    \end{tabular}
    }
    \caption{Comparison of effect of relational graph and also of different components of the relational graph on the overall performance of the model}
    \label{tab:relation_graph_ablation}
\end{table}
In Figure~\ref{fig:relation_qualitative} also analyse the qualitative effect of the relationship graph on the generated caption. We observe that use of the relationship graph allows to better capture the object relationship and thus understand and include semantics of events like waving in the caption.  
\\~\\

\textbf{How much training data is needed?}
We analyze the effect of increase of training data for the task of out-of-context generation. We experiment with different percentages of training data size (w.r.t. to our full data) in Table 4.
\begin{table}[h]
    \centering
    \begin{tabular}{|c|c|c|c|c|}
        \hline 
        Method & Bleu-4 & Cider & Rouge & Meteor \\
        \hline
         Ours (10\%) &  24.6 & 34.04 & 26.49 & 11.94\\
         Ours (20\%) &  28.0 & 43.6 & 30.9 & 13.2\\
         Ours (50\%) &  31.9 & 47.2 & 36.4 & 17.4\\
         Ours (100\%) & 37.4 & 50.1 & 41.5 & 22.4\\
         \hline 
    \end{tabular}
    \caption{Ablation for comparison of different percentages of training data. With respect to training data of 160K images, all context metrics are reported on validation data of 40k images.}
\end{table}

%% file: tex_code/human_evaluation.tex
\section{Human Evaluation}
We conduct a human evaluation to evaluate how convincing model-generated multimedia is to humans. In our human evaluation, we collect 47 responses from people of different demographics and ages on 30 multimedia examples i.e. news caption pairs. Out of which, 15 are real news examples and the rest of the 15 news examples have model-generated captions for which a completely different context is provided manually. We ask the subjects of our study to answer for each of the 30 news examples 2 different questions (1): Do you believe this news is real (Yes/No) and (2) How confident you are in your evaluation (0-5). Note that we specified to each of the subjects that they \textbf{must refrain} from using search engines and only rely on their best judgment to answer these questions. From our 47 respondents, we obtain an average accuracy of 14.83 for the first question for all 30 examples. We also obtain a median score of 15 for the same question for all 30 examples. This allows us to infer that most of the respondents were at best random in guessing which multimedia examples were out-of-context (fake) and which were not (real). In Figure ~\ref{fig:userstudy_qualitative}, 6 out of 15 out-of-context news examples are shown. The figure contains 6 model-generated caption examples denoted as fake respectively along with corresponding real images. We also provide a statistic of how many respondents misclassify each news example. We find that many out-of-context news examples are misclassified as not-out-of-context by the respondents. This demonstrates the hardness of out-of-context multimedia detection task for humans. Thus demonstrating the need for automation in large scale detecting cheap fakes, which would also need automated out-of-context generation (negative examples) to train better discriminative models. 

%\section{Ethical Considerations}
%We are aware of possible misuse of our findings such that given conditional words and an image, the adversary is able to generate out-of-context captions and spread misinformation. However, we suspect that automating this threat is not possible using our model. To automate spread of misinformation using our model on a large scale, adversary needs to build a corpus of real world images and named entities, both of which are plausible. However, to generate malicious captions adversary needs to pair them with a malicious objective, design of which is an open problem. Although we demonstrate that generated out-of-context captions are convincingly real to humans, generating malicious out-of-context captions to spread misinformation on large scale requires additional effort. 

%% file: tex_code/conclusion_ack.tex
\section{Conclusion}
Overall, we show that it is possible to automatically generate and control the semantics of caption for an image, given some conditional word tokens. We present a challenging benchmark to foster the development of defenses against large-scale image re-purposing. From our experimental results, we find that both visual and textual conditional input significantly improve the quality of the generated captions. We also observe that byte-pair encoding helps in improving the captions significantly as it can handle out-of-vocabulary tokens effectively. We also observe that use of a relational graph helps in identifying the underlying object-object relationships. These object relationships enrich the semantics of the caption from the events that could have been ignored by the model.

%% file: tex_code/ethics.tex
\section{Ethical Considerations}
Here, we discuss ethical considerations regarding
our model and proposed task. Cheap fakes are a big source of misinformation online. However, large scale fact checking suffers from lack of high quality annotated training data due to cost and availability issues. Our work is an attempt to proactively bring attention of ML community onto new threats from cheap fakes and proposes a solution that would serve as a baseline for large scale fact checking. We hope that our proposed method serves as a net benefit for society.
\\~\\\textbf{Can this work contribute to large scale misinformation spreading?}
We acknowledge that our method can potentially be misused.
However, we argue that our work cannot be immediately used to generate targeted attacks or spread large scale misinformation. Firstly, since our model is only designed for out-of-context captioning task it cannot be misused to extract an out-of-context image given a caption to suit adversary's narrative. We also argue that to spread misinformation on large scale, an
adversary would have to
manually engineer named entities and events (which is time and money consuming) tailored for their narrative and automatically pair them with out-of-context images of those named entities. We do demonstrate that automatic out-of-context captioning can be convincing to humans. However, additional effort in malicious pairing of named entities and out-of-context images is required for producing misinformation on a large scale from our model.
\\~\\\textbf{What is the generalizability of our claims?} We are expecting our results for out-of-context captioning to primarily apply to western news in the English language. As our model is evaluated on COSMOS dataset ~\cite{cosmos} which contains news image caption pairs in English language that have been source from western media outlets like NewYork Times, Reuters and The
Washington Post.